\documentclass[twocolumn,epsf,psfig]{revtex4-1}
\usepackage{amssymb}
\usepackage{epsfig}
\DeclareMathAlphabet{\pazocal}{OMS}{zplm}{m}{n} 
\usepackage{graphicx}
\usepackage{color}
\usepackage{bm}
\usepackage[normalem]{ulem}     
\usepackage{soul}
\usepackage{amsmath} 
\usepackage{braket}

\begin{document} 
\title{Intrinsic orbital moment and prediction of a large orbital Hall effect in the 
2D transition metal dichalcogenides } 

\author{Sayantika Bhowal} 
\email{bhowals@missouri.edu}
\affiliation{Department of Physics \& Astronomy, University of Missouri, Columbia, MO 65211, USA}
\author{S. Satpathy}
\affiliation{Department of Physics \& Astronomy, University of Missouri, Columbia, MO 65211, USA}
\begin{abstract}
Carrying information using generation and detection of the orbital  current, instead of the spin current, is an emerging field of research,
where the orbital Hall  effect (OHE) is an important ingredient. 
Here, we propose a new mechanism of the OHE that occurs in {\it non-}centrosymmetric materials. We show that the broken inversion symmetry in the 2D transition metal dichalcogenides (TMDCs)
causes a robust orbital moment, 
which flow in different directions due to the opposite Berry curvatures under an applied electric field, leading to a large OHE. This is in complete contrast to the inversion-symmetric systems, where
the orbital moment is induced only by the external electric field.
We show that the valley-orbital locking as well as the OHE both appear even in the absence of the spin-orbit coupling.
The non-zero spin-orbit coupling
leads to the well-known valley-spin locking and the spin Hall effect,
which we find to be  weak, making the TMDCs particularly suitable for direct observation of the OHE,
with potential
application in  {\it orbitronics}.

\end{abstract}

\maketitle 

Orbital Hall effect (OHE) is the phenomenon of transverse flow of orbital 
angular momentum in response to an applied  electric field, similar to the flow of spin angular momentum in the spin Hall effect
(SHE).
 The OHE is more fundamental in the sense that it  occurs with or without the presence of the spin-orbit coupling (SOC), while in presence of the SOC, OHE leads to the additional flow of the spin angular momentum resulting in the SHE.
In fact, the idea of OHE has already been invoked 
to explain the origin of a large anomalous and spin Hall effect in several materials  \cite{Kotani2009,Tanaka,Kotani2008}. Because of this and the fact that OHE is expected to have a larger magnitude than its spin counterpart, 
there is a noticeable interest in developing the  OHE \cite{Go,Go2019,Jo, optically},
with an eye towards future  ``orbitronics" device applications.
%

In this work, we propose a new mechanism of the OHE that  occurs in {\it non}-centrosymmetric materials and explicitly illustrate the ideas for monolayer transition metal dichalcogenides (TMDCs) which constitute the classic example of 2D materials with broken inversion symmetry.
In complete constrast to the centrosymmetric materials \cite{Go, Jo}, where orbital moments are quenched due to symmetry and a non-zero moment develops only due to the symmetry-breaking
applied electric field, here an intrinsic orbital moment is already present in the Brillouin zone (BZ) even without the applied electric field.
Unlike the centrosymmetric systems, the physics here is dominated by the non-zero Berry curvatures, which determines the magnitude of the OHE. 
Our work emphasizes the intrinsic nature of orbital transport  in contrast to the valley Hall effect \cite{Feng, Zhou, Xiao, Xiao2007, ScRep2015}, for example, which can only be achieved by extrinsic means (doping, light illumination, etc.).

 \begin{figure}[t]
\centering
\includegraphics[width=\columnwidth]{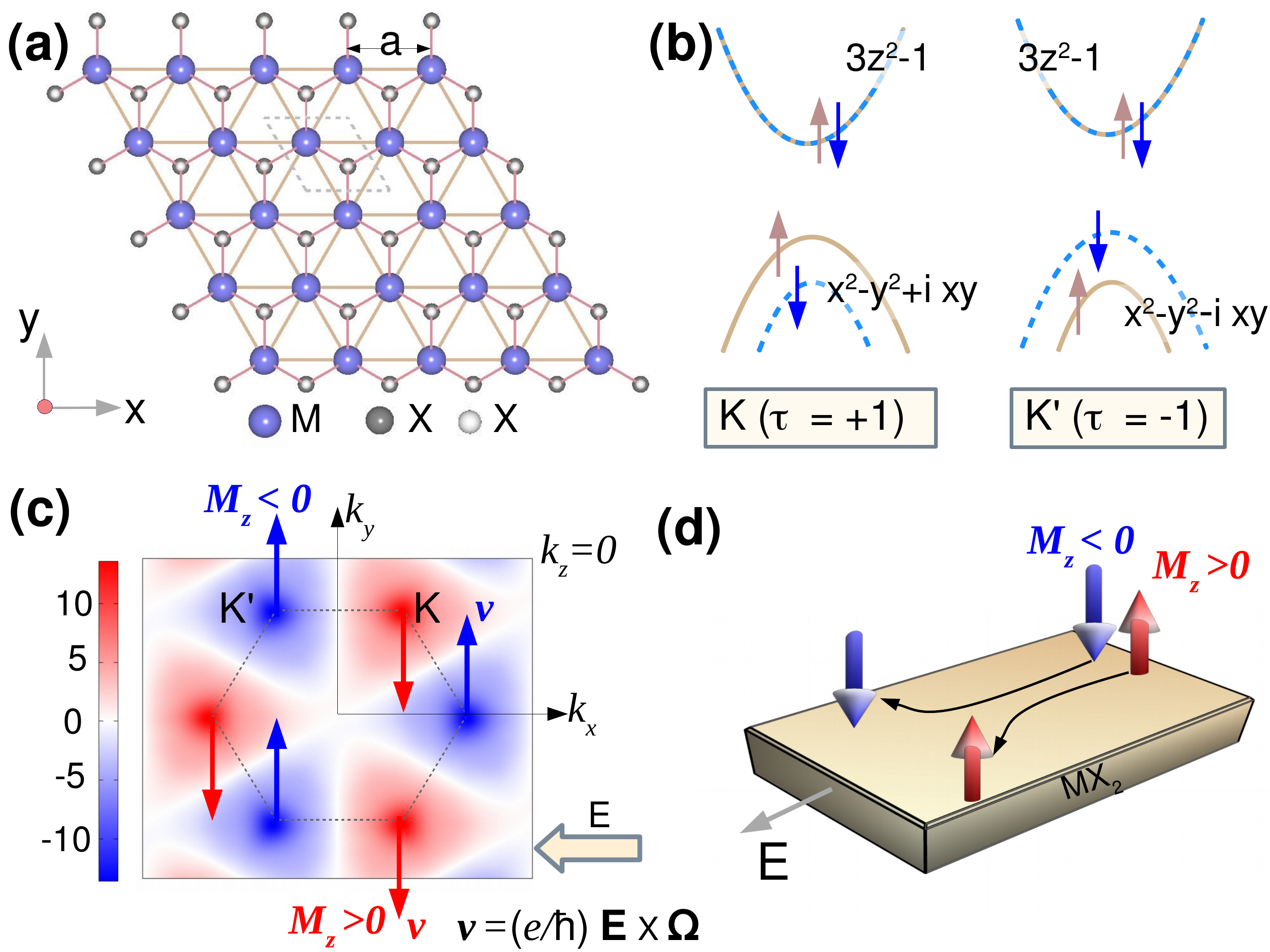}
\caption {Illustration of OHE in monolayer MX$_2$. (a) Crystal structure of  MX$_2$, showing the triangular network of transition metal M atoms as viewed from top. The two out-of-plane chalcogen atoms X occur above and below the plane. (b)  The band structure near $K (-4\pi/3a,0)$ and $ K' (4\pi/3a,0)$, showing the valley dependent spin and orbital characters.  (c) The orbital moment $M_z (\vec k)$ in the BZ and the anomalous velocities ${\bf v}$, indicated by the blue and the red arrows. (d) Orbital moments flow in the transverse direction leading to the OHE.
}
\label{fig1} 
\end{figure}
%



We develop the key physics of the underlying mechanism of the OHE using a tight-binding (TB) model as well as from  
 density-functional calculations.
The effect is demonstrated for  the selected members of the family of 
monolayer TMDCs, viz.,  2H-Mo$X_2$ ($X$ = S, Se, Te), 
where we find a large OHE
and at the same time a negligible intrinsic spin Hall effect, making these materials an excellent platform for the
direct observation of the OHE.



 
%

The basic physics is illustrated in Fig. \ref{fig1}, where we have shown the computed intrinsic orbital moments in the BZ as well as the electron ``anomalous" velocities at the  $K$, $K^\prime$ valleys. Symmetry demands 
that in the presence of inversion (${\cal I}$), orbital moments satisfy the condition
 $\vec M(\vec k) = \vec M(-\vec k)$, while if time-reversal (${\cal T}$) symmetry   is present, we have $\vec M(\vec k) = - \vec M(-\vec k)$. 
 Thus for a non-zero $\vec M(\vec k)$, at least one of the two symmetries must be broken.
 In the present case, broken ${\cal I}$ leads to a nonzero $\vec M (\vec k)$, while its sign changes between the $K$ and $K^\prime$ points due to the presence of $\cal T$. 
The Berry curvatures $ \vec \Omega (\vec k)$ follow the same symmetry properties
 leading to the non-zero anomalous velocity $\vec v  = (e/\hbar) \vec E \times \vec \Omega_{\vec k}$ \cite{Niu} which has opposite directions at the two valleys, and thus leads to the OHE.
These arguments are only suggestive, and one must evaluate
the magnitude of the effect from the calculation of the orbital Berry curvatures \cite{Niu}
as discussed below.

{\it Tight Binding results near the valley points --} 
The valley points ($K$/ $K^\prime$) have the major contributions to the OHE in the TMDCs and this can be studied analytically using a TB model. 
Due to the broken $\cal I$ [see Fig. \ref{fig1} (a)], the chalcogen atoms must be kept along with the transition metal atom (M) in the TB basis set; 
However, their effect may be incorporated via the L\"owdin downfolding \cite{downfolding} producing an effective
TB Hamiltonian for the M-$d$ orbitals with modified Slater-Koster matrix elements \cite {SlaterKoster}.
The effective Hamiltonian, valid near the $K$ and $K^\prime$ valley points reads
\begin{eqnarray}    \label{HKSOC}  
{\cal H } (\vec q ) &=& 
  (\vec d \cdot \vec \sigma )  \otimes I_s   + \frac{\tau \lambda} {2}  (\sigma_z+1) \otimes s_z,
   \end{eqnarray} 
where only terms linear in $\vec q = \vec k - \vec K$ have been kept, 
ignoring thereby the higher-order trigonal warping \cite{Kormanyos}, which are unimportant for the present study.
 Here $\vec s$ and $\vec \sigma$ are respectively the Pauli matrices for the electron spin and the orbital pseudo-spins,
   $ | u \rangle   = (\sqrt{2})^{-1} (|x^2-y^2 \rangle + i\tau |xy \rangle)$ and 
   $|d \rangle
   =|3z^2-r^2\rangle$.
$I_s$  is the $2 \times 2$ identity operator in the electron spin space, $\lambda$ is the SOC constant, and the valley index $\tau = \pm 1$ for the $K$ and $K^\prime$ valleys, 
respectively.
    The TB hopping integrals appear in the parameter $\vec d$, with 
     $d_x = \tau tq_xa, d_y = -tq_ya,$ and $ d_z=-\Delta/2$, where $a$ is the lattice constant,
     $\Delta$ is the energy gap at the $K$ ($K^\prime$) point, and $t$ is an effective inter-band hopping, 
     determined by certain $d-d$ hopping matrix elements.
     We note that Eq. (\ref{HKSOC}) is consistent with  the Hamiltonian derived earlier \cite{Xiao} using the $k\cdot p$ theory.
     The TB derivation has the benefit that it directly expresses the parameters of the Hamiltonian in terms of the specific hopping integrals.

The magnitude of the orbital moment $\vec M (\vec k)$  can be computed for a specific band of the Hamiltonian (\ref{HKSOC}) using
the modern theory  of orbital moment \cite{Niu, Vanderbilt}, viz.,
\begin{eqnarray}\nonumber \label{orb}
 \vec M(\vec k) &=& -2^{-1} ~ \text {Im} [\langle \vec \nabla_k u_{\vec k} | \times ({\cal H} -\varepsilon_{\vec k}) |\vec \nabla_k u_{\vec k} \rangle ] \\ 
& & + ~\text {Im} [\langle \vec \nabla_k u_{\vec k} | \times (\epsilon_F-\varepsilon_{\vec k}) |\vec \nabla_k u_{\vec k} \rangle ],
\end{eqnarray}
where $\varepsilon_{\vec k}$ and $u_{\vec k}  $ are the band energy and the Bloch wave function,
and the two terms in (\ref{orb})  are, respectively, the angular momentum ($\vec r \times \vec v$)
 contribution due to the self-rotation  and due to the motion of the center-of-mass of the Bloch electron wave packet.  
Diagonalizing the $4 \times 4$ Hamiltonian (\ref{HKSOC}),
we  find the energy eigenvalues:
 $\varepsilon^\nu_{\pm} =   2^{-1} [\tau \nu  \lambda \pm ((\Delta - \tau \nu  \lambda)^2 + 4 t^2 a^2 q^2)^{1/2}] $, 
where $\nu = \pm 1$ are the two spin-split states  within the conduction
or valence band manifold, denoted by the subscript $\pm$.
The   wave functions in the basis set 
( $ |u\uparrow \rangle, |d\uparrow \rangle, |u\downarrow \rangle, $ and $ |d\downarrow \rangle$) are
\begin{eqnarray}   \nonumber  \label{wf}
  |u^{\nu = 1}_\pm (q) \rangle &=& {\cal N} \Big[1~~  ( D^\nu \mp    \sqrt {(D^\nu)^2+ d^2}      ) /  d^\nu
  ~~ 0 ~~ 0\Big]^T,   \\  
 |u^{\nu = -1}_\pm (q) \rangle &=& {\cal N} \Big[0 ~~ 0~~ 1 ~~  ( D^\nu \mp    \sqrt {(D^\nu)^2 + d^2}      ) /  d^\nu
 \Big]^T
\end{eqnarray} 
where $ D^\nu = (\Delta  - \nu \tau \lambda)/2 $, $d^\nu  = ta (\tau q_x \pm i \nu q_y)$,  $d^2 = t^2a^2 (q_x^2 + q_y^2)$,
and ${\cal N}$ is the appropriate normalization factor.
With these wave functions, the orbital moments can be evaluated exactly within the TB model from Eq. \ref{orb}. 
For the two valence bands ($\nu = \pm 1$),  the result    is
\begin{subequations}     \label{orbk}         
\begin{gather} 
\hspace{-30mm}
M_z (\vec q) =
 \frac{\tau m_0    D^\nu (D^{-\nu} -\lambda)  \Delta }       {2[(D^\nu)^2 + t^2q^2a^2]^{3/2}}\\
 \approx  \tau m_0 [1+\lambda(3 \nu \tau - 2)/\Delta] (1- 6 \  m_0 q^2 / \Delta),
 \end{gather}
\end{subequations}
where $m_0 =  \Delta^{-1} t^2 a^2 $, only the out-of-plane $\hat z$ component of the orbital moment is non-zero, and the second line is the expansion for small $q $ and $\lambda $, both $ \ll  \Delta$. 


Note the important result (\ref{orbk}) that a large orbital moment $M_z $ exists at the valley points ($\vec q =0$)  and its sign alternates
 between the two valleys ($\tau = \pm 1$) 
({\it valley-orbital locking}).
Furthermore,  it exists even in absence of the SOC ($\lambda = 0$).
For typical parameters,
 $t = 1.22 $ eV, $\Delta =$ 1.66 eV, and $\lambda = 0.08 $ eV, relevant for the monolayer MoS$_2$,
 $m_0 \approx  9.1 $ eV.\AA$^2 \approx  2.4 \mu_B \times (\hbar/e) $. 
 As seen from Eq. \ref{orbk} (b), there is only a weak dependence on $\lambda$.
 
 In fact it is interesting to note that
 the valley-dependent spin splitting [Fig. \ref{fig1} (b)]
  directly follows from the valley orbital moments due to the $\langle \vec L \cdot \vec S  \rangle $ term, which 
  favors anti-alignment of  spin with the orbital moment \cite{Kotani2009}. 
 Thus for the valence bands, the spin-$\downarrow$ band is lower in energy at $K$, while the spin-$\uparrow$ band is lower at $K^\prime$, with a spin splitting of about $ 2 \lambda$. 
 Therefore, the well-known  spin polarization of the bands at the valley points 
 can be thought of to be
 driven by the robust orbital moments via  the perturbative SOC. 
 
 \begin{figure}[t]
\centering
\includegraphics[scale=0.25]{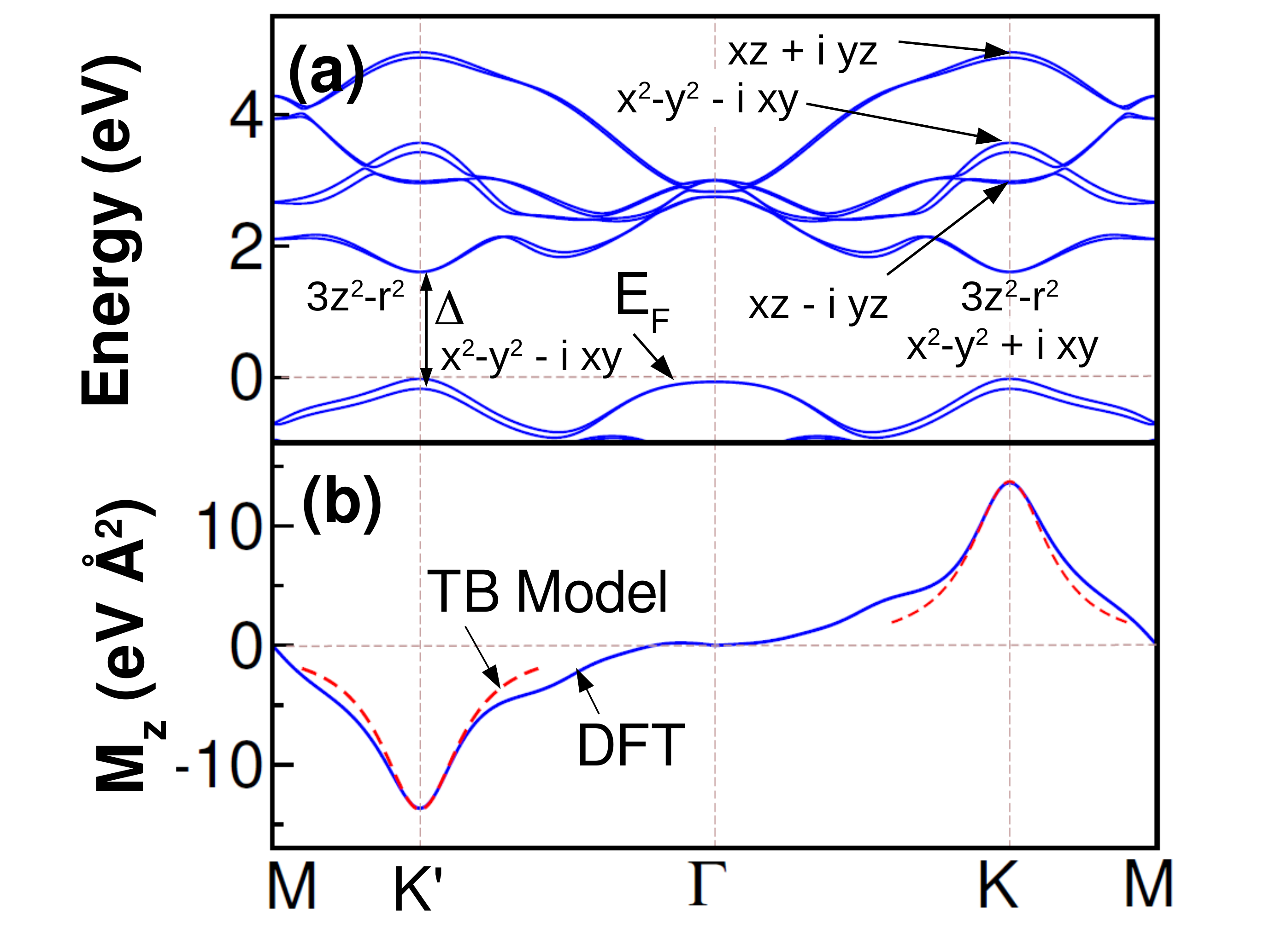}
\caption { (a) Density-functional band structure together with the orbital characters near the valley points and (b) the 
computed sum of the orbital moments ($M_z$) over all occupied bands along selected symmetry lines.
}
\label{fig2} 
\end{figure}

 The orbital moment is the largest at the valley points $K, K^\prime$, as seen from Eq. (\ref{orbk}),  falling off quadratically with momentum $q$. This is also validated by the DFT results shown in Fig. \ref{fig2}.
 The orbital moment at the center of the BZ ($\Gamma$) vanishes exactly due to symmetry reasons, and therefore is expected to be small in the neighborhood of $\Gamma$ as seen from Fig. \ref{fig2} (b) as well.


It is easy to argue that under an applied electric field, the electrons in the two valleys move in opposite directions, so that a net
orbital Hall current is produced.
To see this, we first realize that 
only the Berry curvature term in the  semi-classical expression \cite{Niu} for the electron velocity
$ \dot  {\vec  r}_c = \hbar^{-1}[\vec \nabla_k \varepsilon_{k} +e \vec E \times \vec \Omega (\vec k)]_{\vec k_c}  $
is non-zero  for the two valleys.
Furthermore, only the $\hat z$  component of the Berry curvature survives, which we evaluate near the $K, K^\prime$ valleys  within the TB model using the Kubo formula below.
The result is
\begin{eqnarray}  \label{bck}
&& \Omega^z_n ({\vec  q}) =  - 2 \hbar^2      \sum_{n^\prime \neq n} \frac {{\rm Im} \big[ \langle u_{n{\vec  q}} | v_x | u_{n^\prime{\vec  q}} \rangle  
                        \langle u_{n^\prime{\vec  q}} | v_y | u_{n{\vec  q}} \rangle \big] }
                        {(\varepsilon_{n^\prime \vec q}-\varepsilon_{n  \vec q})^2} \nonumber \\
& & =  \frac{2 M_z (\vec q)}    { \Delta+    \lambda (\nu \tau -2) }
 \approx \frac{2\tau m_0}{\Delta^2} (\Delta +  2 \nu \tau \lambda- 6 m_0 q^2 ). 
\end{eqnarray}
Clearly, $\Omega^z$ has opposite signs for the two valleys, so that $ \vec v \propto \vec E \times \vec \Omega$
is in opposite directions for the $K$ and the $K^\prime$ valley electrons. Thus the positive orbital moment of the $K$ valley moves in one
direction, while the negative orbital moment of $K^\prime$ moves in the opposite direction, leading to a net orbital Hall current.

The magnitude of the orbital Hall conductivity (OHC) may be calculated using the Kubo formula
by  the momentum sum of the orbital Berry curvatures \cite{Go,Jo}, viz., 
\begin{equation}\label{OHC} 
  \sigma^{\gamma,\rm orb}_{\alpha \beta}   =  -\frac{e} { N_k V_c} \sum_{n \vec  k}^{occ} \Omega^{\gamma,\rm orb}_{n,\alpha \beta} ({\vec  k}),
\end{equation}
where $\alpha, \beta, \gamma $ are the cartesian components,
$j^{\rm orb,\gamma}_\alpha = \sigma^{\gamma,\rm orb}_{\alpha\beta} E_\beta$ is the orbital current density along the $\alpha$ direction with the orbital moment along $\gamma$, generated by the electric field along the $\beta$ direction.
  In the 2D systems, $V_c$ is the surface unit cell area, so that
  the conductivity has the dimensions of $(\hbar / e)$ Ohm$^{-1}$.

  The orbital Berry curvature $\Omega^{\gamma,\rm orb}_{n,\alpha\beta}$ in Eq. \ref{OHC} can be evaluated as
\begin{equation} \label{obc}         
 \Omega^{\gamma,\rm orb}_{n,\alpha\beta} ({\vec  k}) = 2 \hbar   \sum_{n^\prime \neq n} \frac {{\rm Im}[ \langle u_{n{\vec  k}} | \mathcal{J}^{\gamma,\rm orb}_\alpha | u_{n^\prime{\vec  k}} \rangle  
                        \langle u_{n^\prime{\vec  k}} | v_\beta | u_{n{\vec  k}} \rangle]} 
                        {(\varepsilon_{n^\prime \vec k}-\varepsilon_{n  \vec k})^2},
\end{equation}
where the orbital current operator is $\mathcal{J}^{\gamma,\rm orb}_\alpha = \frac{1}{2} \{v_\alpha, L_\gamma \}$,  
with $v_{\alpha} =  \frac{1}{\hbar} \frac{\partial H }{ \partial k_\alpha}$
is the velocity operator and $L_\gamma$ is the orbital angular momentum operator.

It turns out that due to the simplicity of the TB Hamiltonian  (\ref{HKSOC}), valid near the valley points,
the orbital and the standard Berry curvatures are the same, apart from a valley-dependent sign, viz., 
\begin{equation} \label{Berry relation}
 \Omega^{z,\rm orb}_{n,yx} (\vec q)  = \tau \times  \Omega^{z}_{n} (\vec q). 
 \end{equation}
 To see this, we take the momentum derivative of (\ref{HKSOC}) to get
\begin{eqnarray}     \label{vx}  
\hbar v_x (\vec q ) &=& 
\left[
{\begin{array}{*{20}c}
  0 & \tau ta & 0 & 0   \\
    \tau ta  & 0 & 0 & 0 \\
   0 & 0 & 0 & \tau ta \\
    0 & 0 & \tau ta  & 0 \\
  \end{array} }  \right] = ta \tau \sigma_x \otimes I_s,
    \end{eqnarray} 
and, 
similarly,
$\hbar v_y (\vec q )   = - ta \sigma_y \otimes I_s$ and $ v_z (\vec q )   = 0$.
Furthermore, in the subspace of the TB Hamiltonian, $L_x = L_y = 0$, and $L_z = \tau \hbar (\sigma_z + 1) \otimes I_s$.
By matrix multiplication, we immediately find that 
$\mathcal{J}^{z,\rm orb}_\alpha = \tau \hbar v_\alpha $ 
and 
$\mathcal{J}^{x,\rm orb}_\alpha = \mathcal{J}^{y,\rm orb}_\alpha = 0$,
which leads to the result (\ref{Berry relation}). 
The expression for the orbital Berry curvature then follows from Eqs. (\ref{bck}) and (\ref{Berry relation}), viz.,
\begin{eqnarray}\label{obck}
 \Omega^{z,\rm orb}_{\nu,yx} (\vec q)  
 =  \frac{2\tau M_z (\vec q)}
 { \Delta+    \lambda (\nu \tau -2) },
 \end{eqnarray}
 where $M_z (\vec q)$ is the orbital moment in Eq. (\ref{orbk}). 
At a general $k$ point, the full expression (\ref{obc}) must be evaluated to obtain the OHC.

This is a key result of the paper, which shows 
 that  the orbital Berry curvatures near the $K$ and $K^\prime$ points are directly proportional to the respective  orbital moments,
 and, more importantly,  they have the same sign at the two valleys as both $\tau$ and $M_z $ change signs simultaneously. 
 Thus, the contributions from these two valleys add up, leading to a non-zero OHC.
 Another important point is that $\Omega^{z,\rm orb}_{\nu,yx}$ exists even {\it without} the SOC, and it has only a weak dependence 
 on $\lambda$ as seen from Eq. (\ref{obck}). 
 Neglecting the $\lambda$ dependence, we see that at both valley points, the contribution to
 the OHC is given by $\Omega^{z,\rm orb}_{\nu,yx} = 2t^2 a^2 / \Delta^2$. 
 In fact, the momentum sum in OHC can be performed analytically in this limit by integrating up to the radius $q_c$ 
 ($ \pi q_c^2 = \Omega_{BZ}$) to yield the result 
 \begin{eqnarray}  \nonumber  \label{sigma-orb}
 \sigma^{z, orb}_{yx}  
 &=&    -\frac{2 e} { (2\pi)^2} \sum_{\nu = \pm 1} \int_0^{q_c} d^2 q  \times \Omega^{z,\rm orb}_{\nu,yx} ({\vec  q}) \\ \nonumber
 &=& \frac{-e} {\pi} \times \Big[ 1- \frac {\Delta}    {\sqrt { \Delta^2 + (32 \pi t^2/\sqrt{3})} } \Big] + O(\lambda^2 /\Delta^2),\\
 \end{eqnarray}
 which is consistent with the anticipated result that  the larger the parameter $t^2/\Delta^2$, 
 the larger is the OHC, primarily because the orbital moment $M_z$ increases.


We pause here to compare the OHE with the related phenomenon of the valley Hall effect, which has been proposed
in the gapped graphene as well as in the TMDCs \cite{Xiao2007, ScRep2015}.
In the valley Hall effect,
 electrons in the two valleys flow in opposite directions,  leading to a charge current and additionally to an orbital current (the valley orbital Hall effect \cite{ScRep2015}), if there is a valley population imbalance (e.g.,  created by shining light).
This is 
in complete contrast to the OHE, which is an {\it intrinsic} effect without any need for population imbalance between the valleys. More interestingly, unlike the valley Hall effect, the OHE described here does not have any net charge current but there exists only a {\it pure} orbital  current. 
Furthermore, in the valley Hall effect, the non-zero valley orbital magnetization \cite{Xiao2007}  explicitly breaks the $\mathcal{T}$ symmetry, which is preserved in the present case.
In this sense the OHE studied here is completely different from the valley Hall effect proposed earlier.

 \begin{figure}[t]
\centering
\includegraphics[width=\columnwidth]{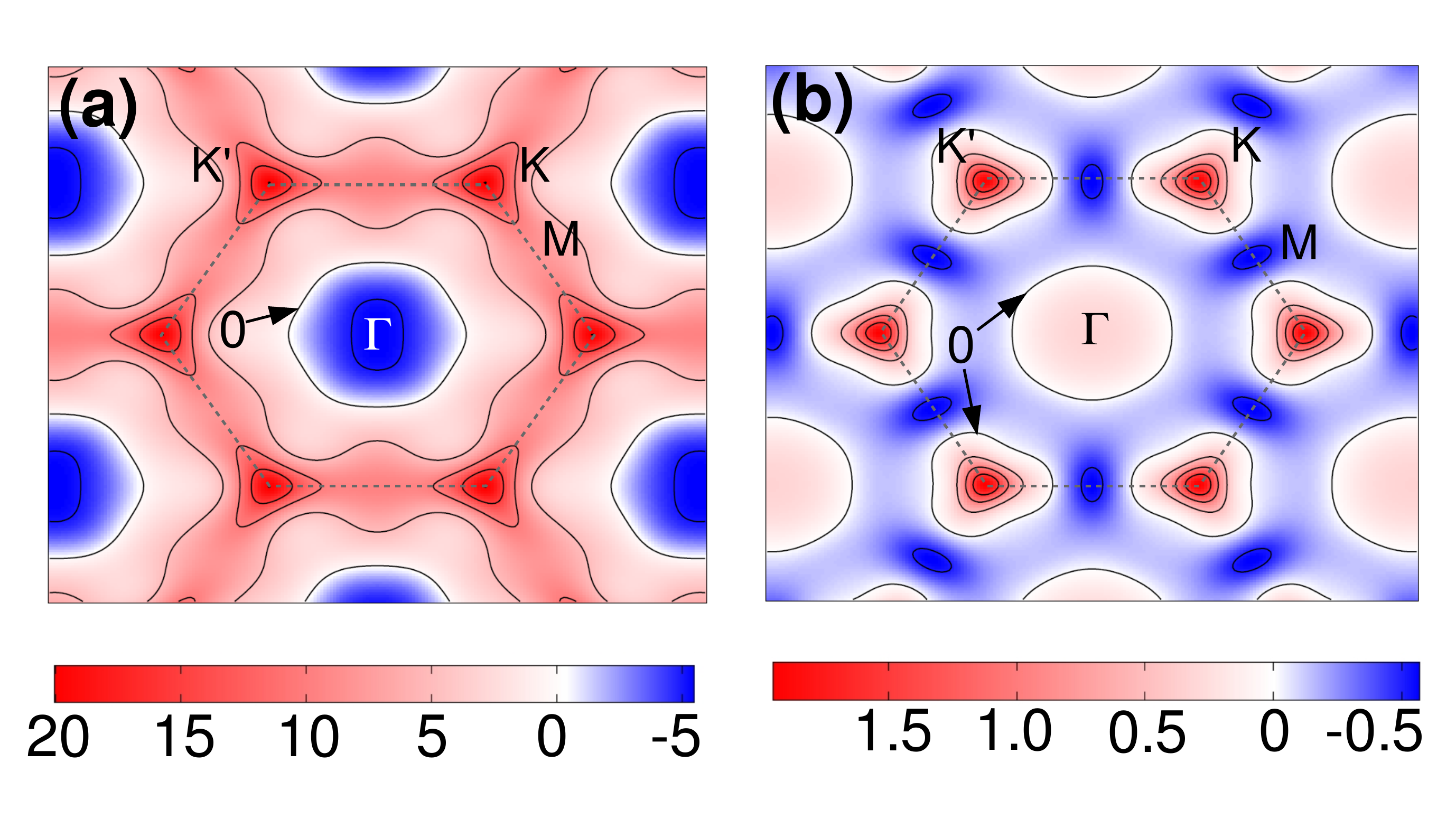}
\caption { (a)
Orbital and (b) spin Berry curvatures (in units of \AA$^2$), summed over the occupied states,
on the $k_z = 0$ plane for 2H-MoS$_2$. The  contours correspond to the tick values on the color bar and the zero contours have been indicated explicitly. 
}
\label{fig3} 
\end{figure}
%

{\it Density functional results} -- We now turn to the DFT results for the monolayer TMDCs. 
Orbital moments were computed using pseudopotential methods \cite{QE} and the Wannier functions as implemented in the Wannier90 code \cite{MLWF, w90} [see Supplementary Materials \cite{SM} for details].
The complementary muffin-tin orbitals based method (NMTO) \cite{NMTO} was used to compute the orbital moment as well as the orbital and the spin Hall conductivities. In the latter method, effective TB hopping matrix elements between the M-$d$ orbitals are obtained for several neighbors, 
which yields the full TB Hamiltonian valid everywhere in the BZ, using which
all quantities of interest are computed. 
The BZ sums for the OHC and spin Hall conductivity (SHC) were computed with $400 \times 400$ $k$ points in the 2D zone.
The computed orbital moments using the Wannier90 or the NMTO method agree quite well.

\begin{table} [b]    
\caption{DFT results for the OHC of the monolayer TMDCs, including  the partial contributions ($\sigma^{z,orb}_{yx} = \sigma_{\rm K} + \sigma_{\rm \Gamma} + \sigma_{\rm rest}$),
$\sigma_{\rm K}$, $\sigma_{\rm \Gamma}$, and $\sigma_{\rm rest}$ being the contributions, respectively, from the valley, $\Gamma$-point, and the remaining regions of the BZ. 
OHC   are
in units of $10^3 \times (\hbar/e)   
 \Omega^{-1}$, while the SHC 
 are in units of $(\hbar/e) \Omega^{-1}$. 
}

\centering
\setlength{\tabcolsep}{7pt}
 \begin{tabular}{c | c c c c  | c}
\hline
 Materials & $\sigma_{\rm K}$ & $\sigma_{\rm \Gamma}$ & $\sigma_{\rm rest}$ & $\sigma^{z,orb}_{yx}$ & $\sigma^{z, spin}_{yx}$\\
\hline
 MoS$_2$ & -9.1 & 1.7 & -3.2 & -10.6 & 1.0\\
MoSe$_2$ & -8.0 & 1.7 & -3 & -9.3 & 1.8\\
MoTe$_2$ &-9.1  & 1.1 & -2.5 & -10.5 & 3.0\\
WTe$_2$ & -8.6 & 1.0 & -2.6 & -10.2 & 9.4 \\
\hline
\end{tabular} 
\label{tab1}
\end{table}

%
The DFT band structure and the corresponding orbital moments  are shown in Fig. \ref{fig1} (c) and 
 Fig. \ref{fig2} for MoS$_2$. As shown in Fig. \ref{fig2} (b), the orbital moments computed from the Hamiltonian (\ref{HKSOC}) near the valley points agree quite well with the DFT results. Note that the total orbital moment (summed over the BZ) vanishes due to the presence
 of ${\cal T}$, though it is non-zero at individual $k$ points.
From the TB model (\ref{HKSOC}), we had studied the orbital moment and the OHE near the valley points.
From the DFT calculations, we can compute the same over the entire BZ, the result of which is shown in Fig. \ref{fig3} (a).
As seen from the figure, the dominant contribution comes from the $k$ space near the valley points $K$, $K^\prime$. 
Since the intrinsic orbital moment near the $\Gamma$ point is absent, the orbital Berry curvature 
in this region takes a non-zero value only due to the orbital moments induced by the applied electric field in the Hall measurement, similar to the centrosymmetric case \cite{Go}.  
This results in a small contribution $\sigma_\Gamma$ to the net OHC, as seen from Table \ref{tab1}, which lists the
partial contributions to the OHC coming from different parts of the BZ. 
Note that there is only one independent component of OHC, viz.,   $\sigma^{z,\rm orb}_{yx}$ = -$\sigma^{z,\rm orb}_{xy}$.

{\it Spin Hall Effect} --
For a material to be a good candidate for the detection of the OHE, the SHC must be small, as both carry angular momentum. 
To this end, we compute the SHC, first from the model Hamiltonian and then from the full DFT calculations.
Analogous to the calculation of the OHC, the SHC can be obtained by the sum of the spin Berry curvatures,
$\Omega^{z,\rm spin}_{\nu, yx} (\vec k)$, evaluated by replacing  the orbital current operator 
with the spin current operator $\mathcal{J}^{\gamma,\rm spin}_\alpha = \frac{1}{4} \{v_\alpha, s_\gamma \}$ in Eq. \ref{obc}.
For the two spin-split valence bands near the valley points in the TB model, we find
\begin{eqnarray}            \label{sbck}
 \Omega^{z,\rm spin}_{\nu, yx} (\vec q) 
=  \frac{ \nu M_z (\vec q)}
 { \Delta+    \lambda (\nu \tau -2) }
 = \frac{\nu\tau}{2}   \Omega^{z,\rm orb}_{yx} (\vec q).
\end{eqnarray} 
Note that $ \Omega^{z,\rm spin}_{\nu, yx} (\vec q)$ has opposite signs for the two spin-split bands and in the limit of
$\lambda = 0$, they exactly cancel everywhere producing a net zero SHC. 
For a non-zero $\lambda$, these two contributions
add up to produce a small net SHC.
Calculating  the contributions from the valley points with a similar procedure as Eq. (\ref{sigma-orb}), we obtain the result 
 $
 \sigma_{yx}^{z, spin} \sim  -e \lambda  (\pi \Delta)^{-1},
$
 in the limit $\lambda \ll \Delta$. This is clearly much smaller than the OHC (\ref{sigma-orb}), by a factor of $\lambda / \Delta$. 
From the DFT results (see Table \ref{tab1}),
we do indeed find that the   SHC is about three orders of magnitude smaller than the OHC.  
Even in doped samples though the SHC is expected to be higher than the undoped sample,
the typical values  \cite{Feng} are nevertheless still an order of magnitude smaller than the computed OHC.
These arguments suggest the TMDCs to be excellent candidates for the observation of OHE, since the intrinsic SHC is negligible in comparison.

In conclusion, we examined the intrinsic OHE in non-centrosymmetric materials and illustrated the ideas for the monolayer TMDCs. 
The broken $\mathcal{I}$ in TMDCs produces a robust momentum-space intrinsic orbital moment $\vec M (\vec k)$, present even in the absence of $\lambda$.
 Due to the opposite Berry curvatures at the valley points $K$ and $K^\prime$, these orbital moments flow in opposite directions, leading to a large OHC
($\approx 10^4~\hbar/e~\Omega^{-1}$).
The vanishingly small intrinsic SHC in these materials make them particularly suitable for the direct observation of the OHE, 
which can be measured by detecting the orbital torque generated by the orbital Hall current \cite{Go2019}. Magneto-optical Kerr effect may also be used to detect the orbital moments accumulated at the edges of the sample due to the OHE \cite{Kato}. Furthermore, the valley-orbital locking can be probed in photon polarized angle-resolved photoemission measurements \cite{Park}.
In addition, it may be possible to tune the OHC by applying a transverse electric field \cite{ Bhowal, ORE}.  
Experimental confirmation of the OHE in the TMDC's may open up new avenues for the realization of orbitronics devices. 

We thank the U.S. Department of Energy, Office of Basic Energy Sciences, Division of Materials Sciences and Engineering for financial support under Grant No. DEFG02-00ER45818.


\begin{thebibliography}{10}


\bibitem{Tanaka}T. Tanaka, H. Kontani, M. Naito, T. Naito, D. S. Hirashima, K. Yamada and J. Inoue, Intrinsic spin Hall effect and orbital Hall effect in 4d and 5d transition metals, Phys. Rev. B {\bf 77}, 165117 (2008).

\bibitem{Kotani2008} H. Kontani, T. Tanaka, D. S. Hirashima, K. Yamada and J. Inoue, Giant Intrinsic Spin and Orbital Hall Effects in Sr$_2$MO$_4$ ($M$ = Ru, Rh, Mo), Phys. Rev. Lett. {\bf 100}, 096601 (2008).

\bibitem{Kotani2009} H. Kontani, T. Tanaka, D. S. Hirashima, K. Yamada and J. Inoue, Giant Orbital Hall Effect in Transition Metals: Origin of Large Spin and Anomalous Hall Effects, Phys. Rev. Lett. {\bf 102}, 016601 (2009).

\bibitem{Go} D. Go, D. Jo, C. Kim, and H.-W. Lee, Intrinsic Spin and Orbital Hall Effects from Orbital Texture, Phys. Rev. Lett. {\bf 121}, 086602 (2018).

\bibitem{Go2019} D. Go and H.-W. Lee, Orbital Torque: Torque Generation by Orbital Current Injection, arXiv:1903.01085 (2019).



\bibitem{Jo} D. Jo, D. Go and H.-W. Lee, Gigantic intrinsic orbital Hall effects in weakly spin-orbit coupled metals, Phys. Rev. B {\bf 98}, 214405 (2018).


\bibitem{optically} V. T. Phong, Z. Addison, S. Ahn, H. Min, R. Agarwal, and E. J. Mele, Optically Controlled Orbitronics on a Triangular Lattice, Phys. Rev. Lett. {\bf  123}, 236403 (2019).


\bibitem{Feng} W. Feng, Y. Yao, W. Zhu, J. Zhou, W. Yao, and D. Xiao, Intrinsic spin Hall effect in monolayers of group-VI dichalcogenides: A first-principles study, Phys. Rev. B {\bf 86}, 165108 (2012).

\bibitem{Zhou} B. T. Zhou, K. Taguchi, Y. Kawaguchi, Y. Tanaka, and K.T. Law, Spin-orbit coupling induced valley Hall effects in
transition-metal dichalcogenides, Communications Physics {\bf 2}, 26 (2019).


\bibitem{Xiao} D. Xiao, G.-B. Liu, W. Feng, X. Xu, and W. Yao, Coupled Spin and Valley Physics in Monolayers of MoS$_2$ and Other Group-VI Dichalcogenides, Phys. Rev. Lett. {\bf 108}, 196802 (2012).

 \bibitem{Xiao2007}  D. Xiao,  W. Yao, and Q. Niu, Valley-Contrasting Physics in Graphene: Magnetic Moment and Topological Transport, Phys. Rev. Lett. {\bf 99}, 236809 (2007).
 
\bibitem{ScRep2015} Z. Song, R. Quhe, S. Liu, Y. Li, J. Feng, Y. Yang, J. Lu, and J. Yang, Tunable Valley Polarization and
Valley Orbital Magnetic Moment
Hall Effect in Honeycomb Systems
with Broken Inversion Symmetry, Sc. Rep. {\bf 5}, 13906 (2015).

\bibitem{Niu} D. Xiao, J. Shi and Q. Niu, Berry Phase Correction to Electron Density of States in Solids, Phys. Rev. Lett. {\bf 95}, 137204 (2005); D. Xiao, M.-C. Chang, and Q. Niu, Berry phase effects on electronic properties, Rev. Mod. Phys. {\bf 82}, 1959 (2010). 

\bibitem{downfolding} P.-O. L\"owdin, A Note on the Quantum‐Mechanical Perturbation Theory, J. Chem. Phys. {\bf 19}, 1396 (1951).

\bibitem{SlaterKoster} J. C. Slater and G. F. Koster, Simplified LCAO Method for the Periodic Potential Problem, Phys. Rev. {\bf 94}, 1498 (1954).







\bibitem{Kormanyos} A. Korm\'anyos, V. Z\'olyomi, N. D. Drummond, P. Rakyta, G. Burkard, and V. I. Fal'ko, Monolayer MoS$_2$: Trigonal warping, the $\Gamma$ valley, and spin-orbit coupling effects, Phys. Rev. B {\bf 88}, 045416 (2013).

\bibitem{Vanderbilt} D. Ceresoli, T. Thonhauser, D. Vanderbilt, and R. Resta, Orbital magnetization in crystalline solids: Multi-band insulators, Chern insulators, and metals,  Phys. Rev. B {\bf 74}, 024408 (2006).







\bibitem{QE} Giannozzi, P. {\it  et al.} QUANTUM ESPRESSO: a modular and open-source software project for quantum simulations of materials, {\it J. Phys. Condens. Matter} {\bf 21}, 395502 (2009).

\bibitem{MLWF} Marzari, N. \& Vanderbilt,  D. Maximally localized generalized Wannier functions for composite energy bands, 
{\it Phys. Rev. B} {\bf 56}, 12847 (1997); I. Souza, N. Marzari and D. Vanderbilt, Maximally Localized Wannier Functions for Entangled Energy Bands, Phys. Rev. B {\bf 65}, 035109 (2001).

\bibitem{w90} M. G. Lopez, D. Vanderbilt, T. Thonhauser, and I. Souza Wannier-based calculation of the orbital magnetization in crystals, Phys. Rev. B {\bf 85}, 014435 (2012); A. A. Mostofi, J. R. Yates, G. Pizzi, Y. S. Lee, I. Souza, D. Vanderbilt and N. Marzari, An updated version of Wannier90: A Tool for Obtaining Maximally Localised Wannier Functions, Comput. Phys. Commun. {\bf 185}, 2309 (2014).

\bibitem{SM} Supplementary Materials describing the density functional methods and additional DFT results. 

\bibitem{NMTO} O. K. Andersen and T. Saha-Dasgupta, Muffin-tin orbitals of arbitrary order, Phys. Rev. B {\bf 62}, R16219 (2000).









\bibitem{Kato} Y. K. Kato, R. C. Myers, A. C. Gossard, and D. D.
Awschalom, Intrinsic Spin and Orbital Hall Effects from Orbital Texture, Science 306, 1910 (2004).

\bibitem{Park} S. R. Park, J. Han, C. Kim, Y. Y. Koh, C. Kim, H. Lee,
H. J. Choi, J. H. Han, K. D. Lee, N. J. Hur, M. Arita, K. Shimada,
H. Namatame, and M. Taniguchi, Chiral Orbital-Angular Momentum in the Surface States of Bi$_2$Se$_3$,  Phys. Rev. Lett. {\bf 108}, 046805 (2012).

\bibitem{Bhowal} S. Bhowal and S. Satpathy, Electric field tuning of the anomalous Hall effect at oxide interfaces, npj Computational Materials {\bf 5} (1), 61 (2019).

\bibitem{ORE} D. Go, J.-P. Hanke, P. M. Buhl, F. Freimuth, G. Bihlmayer, H.-W. Lee, Y. Mokrousov and Stefan Bl\"ugel, Toward surface orbitronics: giant orbital magnetism from the orbital
Rashba effect at the surface of $sp$ metals, Sci. Rep. {\bf 7}, 46742 (2017).


\end{thebibliography}
\end{document}